\documentclass[12pt]{iopart}
\usepackage{graphicx}
\usepackage{amsfonts}
\usepackage{amssymb}
\begin{document}

\title{Different steady states for spin currents in noncollinear multilayers}

\author{Peter M Levy and Jianwei Zhang}
\address{Department of Physics, New York University, New York, NY,10003,USA}

\ \ \  \ \  \ \ \ \  \ \  \ \today   

\pacs{72.25.-b,72.15.Gd, 73.23.-b}

\begin{abstract}
We find there are at least two different steady states for transport across
noncollinear magnetic multilayers. In the conventional one there is a
discontinuity in the spin current across the interfaces which has been
identified as the source of current induced magnetic reversal; in the one
advocated herein the spin torque arises from the spin accumulation transverse
to the magnetization of a magnetic layer. These two states have quite
different attributes which should be discerned by current experiments.
\end{abstract}

\maketitle

Recent interest has focused on electron transport across metallic multilayers
which contain the 3d transition metals, especially the spin torque acting on a
magnetic layer as current is driven across noncollinear magnetic
multilayers.\cite{slon1} To calculate the transport properties of multilayered
structures one would ideally calculate the wavefunctions for the entire
structure and then evaluate the conductivity in the presence of impurities;
\cite{kuising-waintal} in this approach one retains the long range nature of
the conductivity, which is important even for diffusive metals.\cite{kane}
However what is usually done is to adopt a layer-by-layer approach in which
one solves for the transport within homogeneous layers and connects the out of
equilibrium distribution functions with transmission and reflection
coefficients at the interfaces. These functions are statistical density
matrices and contain less information than wavefunctions, therefore one should
anticipate some loss of information that may alter the conductivity one
calculates for the entire layered structure. Indeed this has been found when
discussing giant magnetoresistance, and it was shown that it necessary to
include either spin dependent electrochemical potentials or equivalently spin
accumulation in the layer-by-layer approach which one would not explicitly do
if one determined the wavefunctions for the entire approach.\cite{our work}
Most of the analyses of transport in multilayers has been done on collinear
structures and as far as one can tell the use of spin dependent electric
fields or accumulation has been able to make up for the putative shortcomings
of the layer-by layer approach for the conductivity of the whole structure.
However when one studies noncollinear magnetic multilayers the previous
constructs used to shore up the layer-by-layer approach may well be
insufficient and, unless augmented by additional artifices, can lead to
erroneous predictions, e.g., as to the microscopic origin of spin torque. Here
we present the case for adding such a new construct: current induced
contributions to the transmission coefficients which connect the out of
equilibrium distribution functions between layers.

When current is driven across a magnetically multilayered structure the first
response of the itinerant carriers is a rapid accumulation of charge around
the interfaces to establish a steady state charge current, i.e., one that does
not vary with time; in metallic structures this accumulation is confined to a
screening length of the order of angstr\"{o}ms. The difference in the density
of states at the Fermi level and scattering rates for spin up/down electrons
leads to a current which is spin polarized, i.e., $j_{\uparrow}-j_{\downarrow
}\neq0$ in the magnetic layers, while the current in the normal layers is not
polarized. This discontinuity in the polarized current at interfaces between
ferromagnetic and normal metal layers (N/F) is the \textit{source} of spin
accumulation about the interface; its length scale, known as the spin
diffusion length, is controlled by the rate of spin flip $\tau_{sf}^{-1}$ in
the layers; it is only after one has established this spin accumulation that
the spin polarized current from one magnetic layer can be transmitted to
another provided that the normal (non magnetic) spacer layer is less than the
spin diffusion length.\cite{shufeng} It is also true that one achieves a
steady state spin current after $t\gtrsim\tau_{sf}$ which is of the order of a
picosecond, say in Cu; conversely if one does not consider the source term at
interfaces there is no accumulation and no steady state in the thermodynamic
sense, i.e., one cannot achieve a state of maximum entropy
production.\cite{ziman}

For collinear structures when the spin accumulation created at one N/F
interface arrives at the other interface of the normal layer it is
superimposed on the polarization that exists in equilibrium; while it alters
the interface scattering potential at the second interface it does not
contribute in linear response, rather it is a nonlinear effect. It is for this
reason that one uses the transmission coefficients found from equilibrium band
structure calculations. However for noncollinear structures the spin
polarization that is superimposed has a component that is transverse to the
polarization of the second interface. This does not lead to a current driven
repopulation of equilibrium states ,i.e., additional (nonlinear) accumulation,
but to a \textquotedblleft current induced spin flip\textquotedblright%
\ between states of opposite spin; this has been recently shown to lead to a
new contribution to the transmission coefficient that enters in linear
response.\cite{levy} We have found that this additional term makes the
difference in the predictions based on models which use realistic band
structures for the individual layers and equilibrium transmission coefficients
to connect the nonequilibrium distribution functions across
interfaces,\cite{discontinuities} and those which use free electron bands
which maintain spin coherence across the layers and mimic the scattering at
interfaces by phenomenological interface resistances,\cite{asya} i.e., when
one includes the additional interface scattering in the former models one
retrieves the features found in the latter.

The rationale behind the conventional treatment of current induced spin torque
is the following. The transmission of the spin current across the normal
metal/ferromagnetic (N/F) interfaces is described by spin dependent reflection
and transmission scattering amplitudes, and the Stoner model of spin split
bands for the magnetic layers has provided a good description of electron
transport in collinear magnetic multilayers and provides an explanation for
\textit{one} of the origins of giant magnetoresistance, spin dependent
interface scattering due to differences in band structure at the N/F
interfaces; the model does not involve spin flips and indeed provides a basis
for understanding why spin-flips, which are high energy excitations, are rare
( impurity induced). This same model has been applied to noncollinear
structures in which the angle between the magnetic layers is different from
0$^{0}$ and 180$^{0}$. One again with no spin flip scattering at the
interfaces and by using the same parameters for the spin dependent interface
scattering one finds the spin currents that were continuous in the collinear
case now have discontinuities at the interfaces.\cite{discontinuities} Indeed
in this description these discontinuities represent the transfer of spin
angular momentum from the current to the magnetic layer and thereby create a
\textquotedblleft spin torque\textquotedblright\ which eventually leads to
current induced magnetization reversal.\cite{slon1} The origin of the
discontinuity in the spin current can be traced back to the transmission and
reflection coefficients at N/F interfaces. In this view the spin angular
momentum lost by the spin current goes to the background magnetization which
implies that a steady state for the spin current is not achieved before the
background moves in such a manner as to remove the discontinuity in the spin
current, which is of the order of a nanosecond.

An alternate view has been proposed in which a steady state spin current is
achieved on the time scale of the longitudinal spin flip time, which is of the
order of a picosecond.\cite{asya,prl} The salient difference in the two
views is that we find that the discontinuity in the spin current at the
interfaces \textit{drives} a transverse spin accumulation which in turn
achieves a steady state distribution for the spin current; in this manner the
discontinuity is counterbalanced by accumulation so that $\partial
_{t}f(k,r,t)=0$. In the conventional treatments one has not considered this
transverse accumulation, rather one solved the transport equations of motion
by setting $\partial_{t}f=0$, and thereby assumed steady state; if one
inquires about the time to achieve this solution one would come up with a
nanosecond. We have solved the\textit{ time dependent} equations of motion for
transport across a noncollinear F/N/F structure using the same method we used
for collinear structures;\cite{shufeng} the new ingredient is the transverse
components of the discontinuity in the spin current. For thin normal
layers,$\thicksim10nm$ , it takes of the order of several femtoseconds for the
spin accumulation created at one N/F interface to reach the other interface.
When the layers are noncollinear the longitudinal accumulation created by the
first interface has a component transverse to the magnetization of the second
magnetic layer. We find this drives a transverse accumulation of spin in the
second magnetic layer which leads to a steady state solution in which the spin
current is continuous across the interfaces.\cite{jianwei3} In the presence of
differences in band structure across interfaces it is necessary to include the
current driven spin flip scattering at interfaces to inject a spin current
with a component transverse to the magnetization of a ferromagnetic layer; if
one limits oneself to the equilibrium transmission coefficients there is no
excitation of the transverse spin current across an interface and thereby no
transverse accumulation, i.e., the steady state solutions with and without the
interface spin flip scattering are indeed different. For example we have found
the resistivity for noncollinear structures is always lower when we consider
this additional scattering.\cite{jianwei2}

To exhibit the main differences in our approach compared to the conventional
one we have solved for electron transport in a noncollinear magnetic trilayer
F/N/F by using the semiclassical Boltzmann equation of motion; we assumed that
the nonmagnetic spacer is thin compared to the spin diffusion length so that
one can effectively reduce the transport calculation that of two noncollinear
ferromagnetic layers with transmission coefficients given in terms of the N/F
transmission amplitudes of the trilayer.\cite{bauer} We take spin split but
otherwise free electron bands; this is sufficient to model the band mismatch
in the 3d transition-metal ferromagnets. Parenthetically, we have derived the
equations for a Fermi sea of electrons; when we neglect the current driven
corrections to the Fermi sea and focus only on the Fermi surface (see below)
our equations are the same as those we find by using the $s-d$
model.\cite{prl} In each layer the energy and density vary slowly on the
length scale of the Fermi wavelength so that we can limit ourselves to the
first term in the gradient expansion of the equation of motion.\cite{rammer
and smith} When we limit ourselves to linear response in the external electric
field we f{}ind the equations of motion for the elements of the spinor density
matrix for each momentum state on the Fermi surface $k_{p}$ are,\cite{jianwei}%

\begin{equation}
\partial_{t}f_{p}+v_{p}^{x}\partial_{x}f_{p}-eEv_{p}\delta(\varepsilon
-\varepsilon_{F})=-\frac{f_{p}-\left\langle f_{p}\right\rangle }{\tau_{p}%
}-\frac{f_{p}-\left\langle f_{p^{\prime}}\right\rangle }{\tau_{sf}},\label{d}%
\end{equation}
and
\begin{equation}
\partial_{t}f_{p}^{\pm}+v_{p}^{x}\partial_{x}f_{p}^{\pm}\mp i\frac{J_{p}%
}{\hbar}f_{p}^{\pm}=-\frac{f_{p}^{\pm}-\left\langle f_{p}^{\pm}\right\rangle
}{\tau_{p}},\label{e}%
\end{equation}
where we have used a simplified index $p$ to denote the momentum $k_{p}$ of a
state on the $n^{th}$ sheet of the Fermi surface(we suppress this index),
$p^{\prime}$ are states of opposite spin to $p$, the average $\left\langle
f_{p}\right\rangle $ represents elastic scattering to all states on the Fermi
surface, $v_{p}^{x}$ is the component of the Fermi velocity along the electric
field $E$, $J$ is the magnetic part of the energy and we have limited the
current induced variations of the distribution function to those along the
growth direction of a multilayered structure $x$ (also the field direction).
The diagonal elements $f_{p}=f_{\alpha\alpha}(k_{p},x)$ represents the
occupancy of the state $k_{p}$; in equilibrium it is given by the Fermi
function so that only the spin state $\alpha$ that crosses the Fermi level is
occupied while the other is zero and we do not consider it further. The off
diagonal elements $f_{p}^{\pm}\sim f_{\uparrow\downarrow}(k_{p},x)$, which we
call a current induced spin coherence,\cite{coherence} represent coherences
between the state $k_{p}$ on the Fermi surface and the states with opposite
spin; these coherences occur when we drive a spin current across a N/F
interface.\cite{levy} The scattering terms include those for states of the
same spin on the Fermi surface $\tau_{p}^{-1}$ as well as those between sheets
of opposite spin $\tau_{sf}^{-1}$; as $\tau_{sf}^{-1}\ll$ $\tau_{p}^{-1}$ we
include the latter only to have well defined boundary conditions on our
distribution functions.

The steady state ($\partial_{t}f_{p}=0$) solutions for the longitudinal
components $f_{p}$ are well know.\cite{valet-fert} From Eqs.(\ref{d}) and
(\ref{e}) we see that the electric f{}ield only creates out of equilibrium
longitudinal components of the distribution functions; in a homogeneous
magnetic layer there is no coupling to the transverse components $f_{p}^{\pm}%
$. However when the spin current from one layer is injected into another
noncollinear magnetic layer the transverse components can be excited
\textit{provided }one includes the current induced spin flip scattering at the
interface; their inclusion removes the discontinuity in the spin current at
the interfaces. When we neglect collisions (the rhs of Eq.(\ref{e})) the
transverse solutions in steady state are $f_{p}^{\pm}(x)\sim\exp\pm
\mathbf{i}(J_{p}/\hbar v_{p}^{x})x$; when we average this over the Fermi
surface we f{}ind an interference between individual $p$ or $\mathbf{k}$
states so that the transverse components of the spin currents in the magnetic
layers, $j_{x}^{\pm}(x)\sim\int v_{x}(k)f^{\pm}(k,x)dk$ , can be f{}it to a
form \textit{approximating} an exponential decay $\sim\exp-x/\lambda_{tr}$ .
In the ballistic regime $\lambda_{tr}=d_{J}\equiv h\overline{v_{F}/J}$ where
the bar denotes an average over states on the $n^{th}$ sheet of the Fermi
surface under consideration; \cite{jianwei} while for diffusive systems where
we consider the collision terms we find $\lambda_{tr}=\lambda_{J}\equiv
\sqrt{d_{J}\lambda_{mfp}/3\pi}$ when using the spin diffusion
equation.\cite{prl} Therefore when components of the spin current are injected
into a magnetic layer that are transverse to its magnetization, we find they
propagate a distance $\lambda_{tr}$ before decaying; as this distance is an
order of magnitude greater than the Fermi wavelength one can describe the
transverse spin currents in the semiclassical Boltzmann approach. As
$\lambda_{tr}$ is comparable to the thickness of the magnetic layers
undergoing switching one cannot assume the transverse components of spin
currents are entirely absorbed in such thin layers.

The salient results we find from the steady state solutions of the above
equations are: transverse spin current and accumulation exist in the magnetic
layers up to $\lambda_{tr}\thicksim$3 nm from the interface in the ballistic
regime, there is no discontinuity in the spin current provided one has
included the spin flip at the interface, and when we include diffuse
scattering in the layers $\lambda_{tr}$ is by and large different from the
$\lambda_{J}$ found from the spin diffusion equation.\cite{asya}
\begin{figure}
\begin{center}
\includegraphics[height=6.2cm]{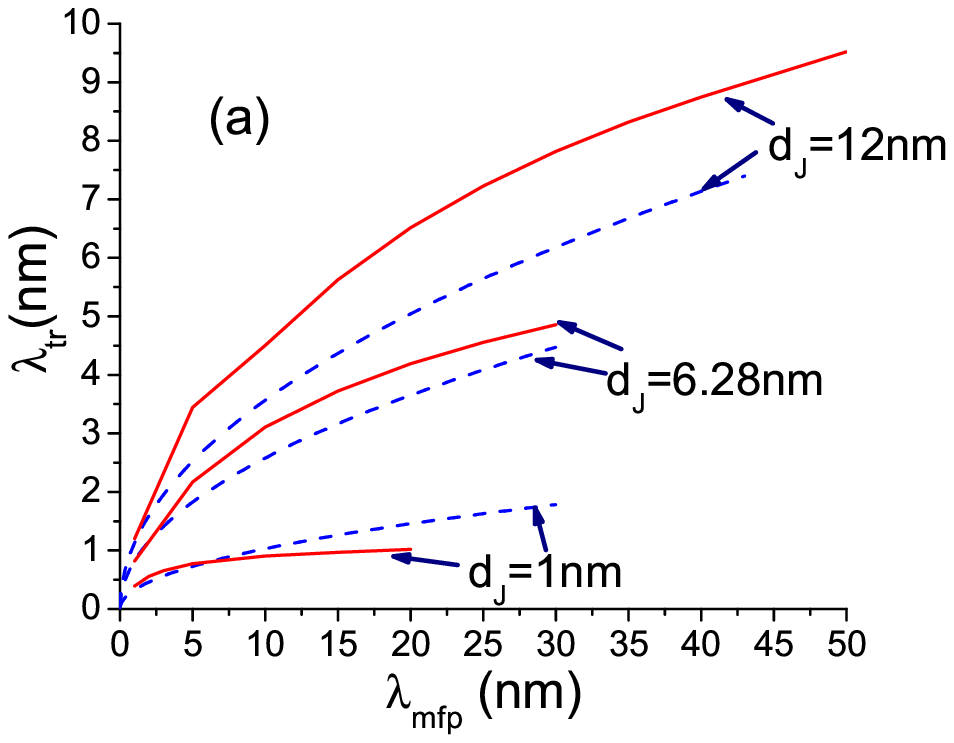}
\includegraphics[height=6.2cm]{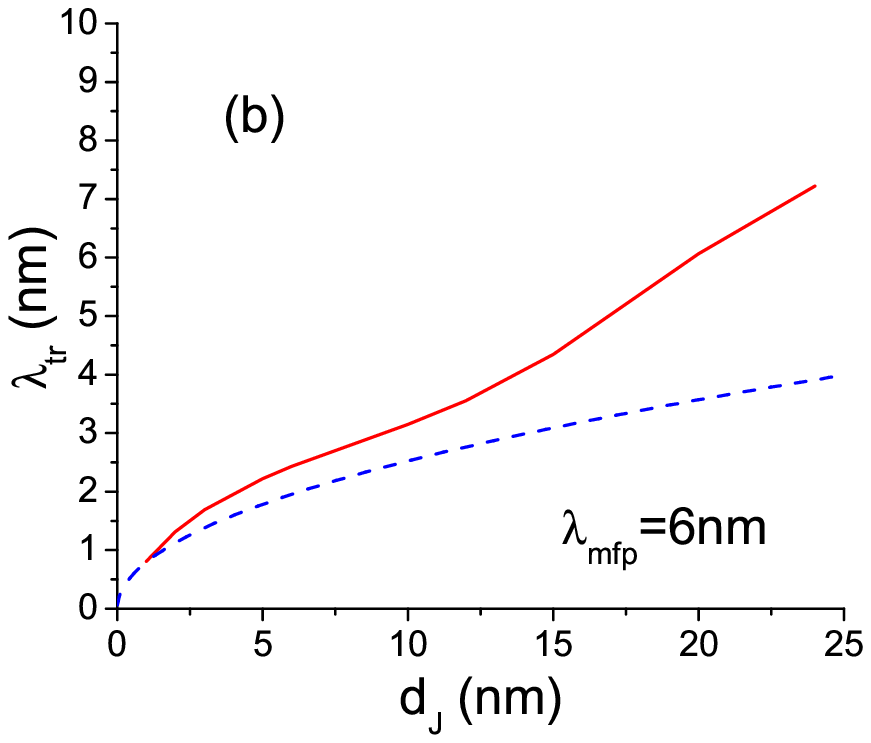}
\caption{The transverse length of spin current $\lambda_{tr}$.
The solid lines are the Boltzmann result, the dash lines are the diffusion result.
In fig(a) we show the variation as a function of mean free path for different $d_J$;
in (b) the variation with $d_J$ for a mean free path equal to $6$nm.}
\end{center}
\label{fig01}
\end{figure}
In Fig.1 we show the variation of the transverse decay length $\lambda_{tr}$ as a function
of the exchange splitting $J$ and the mean free path $\lambda_{mfp}$; for the
parameters that describe $Co$ we find the solutions found from the Boltzmann
equation are reasonably well described by the spin diffusion equation in which
one has neglected the spin splitting of the bands but included diffuse
interface scattering as a resistance.\cite{asya} In these calculations the
coherence between states of opposite spin are kept intact, because we have not
used a spin polarized effective one electron description of the bands in the
ferromagnetic layers. When we use the spin split band description we are able
to retrieve the hallmark of the coherence, i.e., continuity of spin current
across the interface, if we include the spin flip scattering coefficients at
the interfaces; by using only the equilibrium transmission coefficients we
find the spin current is discontinuous. We conclude that for spin transport in
noncollinear multilayers it is more important to keep track of the spin
coherence between states of opposite spin in ferromagnetic layers than the
precise details of the spin split band structure.

We have also started to determine the time dependent solutions for the
transport equations in noncollinear structures.\cite{jianwei3} Initially there
is a discontinuity in the spin current at the interfaces; the longitudinal
component of this discontinuity is relaxed by the random spin flip scattering
in the bulk of the layers.\cite{shufeng} The transverse components are relaxed
by the third term on the left hand side of Eq.(\ref{e}); however if one does
not include the spin flip scattering at the interfaces the transverse
components of the distribution function are not excited and there is no
relaxation, i.e., one does not have a steady state solution, rather the
discontinuity in the transverse component of the spin current remains until
the free magnetic layer switches. We have also calculated the resistance of
noncollinear trilayers (modelled as bilayers), and always find their
resistance is \textit{lower }when we include the interface spin flip
scattering.\cite{jianwei2}

In conclusion the steady state arrived at by using only the equilibrium
transmission coefficients for noncollinear magnetic multilayers has different
attributes from the one reached when either one uses the current induced
interface scattering, or neglects the spin splitting of the band structure.
They differ in their resistances, length scales of spin transfer, magnitudes
of spin torque created, and the time to reach steady state. Specifically,
providing all other things are equal, the spin torque in our model is less
than that found in the conventional ones when the thickness of the layer to be
switched is less than the transverse decay length which is of the order of 2-3
nm. These differences should be within the reach of current experimental
conditions to ascertain which picture better fits the data.

We would like to thank Professor Shufeng Zhang for very helpful discussions.
Research supported by the National Science Foundation under Grant DMR 0131883.


\begin{thebibliography}{99}
\bibitem {coherence}The off diagonal elements of a density matrix $|\phi
_{a}><\phi_{b}|$ refer to coherences between states; only the diagonal
elements $a=b$ represent the occupancy of a state.here we are interested in
those induced by currents across noncollinear magnetic structures at the Fermi surface.
\end{thebibliography}
\end{document}